\documentclass[%
reprint,
aps,
pra
]{revtex4-2}

\usepackage[usenames]{color}
\usepackage{colortbl}
\usepackage{graphicx}
\usepackage{dcolumn}
\usepackage{bm}
\usepackage{amsmath}
\usepackage{amssymb}
\usepackage[utf8]{inputenc}

\usepackage{comment}
\usepackage{dsfont}
\usepackage{xcolor}
\usepackage{multirow}

\begin{document}

\preprint{APS/123-QED}

\title{
Second-law-allowed temporal cooling of the coldest reservoir without external refrigeration 
}

\author{I. V. Vovchenko$^{1,3}$, A. A. Zyablovsky$^{1,2,3}$, A. A. Pukhov$^{1,2}$, E. S. Andrianov$^{1,2,3}$}
\affiliation{%
 $^1$Moscow Institute of Physics and Technology, 9 Institutskiy pereulok, Dolgoprudny 141701, Moscow region, Russia;
}%
\affiliation{
 $^2$Institute for Theoretical and Applied Electromagnetics, 13 Izhorskaya, Moscow 125412, Russia;
}
\affiliation{%
 $^3$Dukhov Research Institute of Automatics (VNIIA), 22 Sushchevskaya, Moscow 127030, Russia;
}%

\date{\today}

\begin{abstract}
Non-equilibrium quantum thermodynamics is an intensively developing field with many existing applications.
We study the dynamics of temperatures and chemical potentials of fermionic reservoirs coupled to an open quantum system.
We show that heat transfer from the coldest reservoir to the hottest one is allowed by the Clausius inequality and results in transient cooling of the coldest reservoir without additional external refrigeration.
We show that during the establishment of thermal and chemical equilibrium, non-monotone evolution of reservoirs' temperatures and chemical potentials is possible, including changes in reservoirs' temperatures and chemical potentials orderliness.
Achieved results can be used in the design of quantum thermal machines and nanoelectronic devices.
\end{abstract}

\maketitle




Non-equilibrium quantum thermodynamics is a rapidly developing field due to many applications in quantum transport devices \cite{pekola2021colloquium,kohler2005driven,kato2015quantum,ronzani2018tunable,smith1996low,taylor2001ab,gordeeva2020record,islam2024evaluating}, phonon and photon logic elements \cite{jiang2015phonon,li2012colloquium,malavazi2024detuning,ojanen2008mesoscopic}, quantum thermal machines \cite{binder2018thermodynamics,bhattacharjee2021quantum,leggio2015quantum,gemmer2009quantum,kosloff2014quantum,uzdin2015equivalence}, devices for thermodynamic computing \cite{fry2017physical,boyd2022shortcuts,hylton2020thermodynamic,aifer2024thermodynamic} and entanglement creation \cite{al2008sudden,vznidarivc2012entanglement,arnesen2001natural,wang2001entanglement,orszag2010coherence,scala2008dissipation,scala2011robust,vovcenko2021dephasing}.
Contrary to classical thermodynamics, which exploits an axiomatic approach to macroscopic thermodynamic values \cite{adkins1983equilibrium,kondepudi2014modern,kestin1979course,deffner2019quantum,swendsen2017thermodynamics}, quantum thermodynamics adopts methods from the theory of open quantum systems (OQS), obtaining macroscopic thermodynamic values via analysis of operators' dynamics \cite{potts2024quantum,binder2018thermodynamics,alicki2018introduction,gemmer2009quantum,deffner2019quantum,lindblad2001non}.
Accomplishment of classical thermodynamics laws \cite{kosloff2013quantum,spohn1978entropy,levy2014local,bera2017generalized} as well as classical definition of heat flows \cite{lipka2024fundamental,sapienza2019correlations,bera2017generalized,alipour2016correlations,pyharanta2022correlation} are questionable, and both should be considered in accordance with microscopic quantum description.
For example, zero and second laws of thermodynamics in classical form can be fulfilled or broken \cite{spohn1978entropy,spohn1973irreversible,levy2014local,shishkov2020perturbation,trushechkin2021unified,trushechkin2016perturbative}.
Notably, quantum properties strongly influence heat flows \cite{binder2018thermodynamics,pekola2015towards,martensen2019transmission,topp2015steady,lipka2024fundamental,sapienza2019correlations,bera2017generalized,alipour2016correlations,pyharanta2022correlation,prech2023entanglement}.

The second law of thermodynamics (SLoT) is governing relaxation of mesoscopic and macroscopic systems to equilibrium state \cite{grandy2008entropy,Landau_StatPhys,deffner2019quantum,zaitsev2008introduction,lieb2003entropy,lieb1999physics} prohibiting entropy decrease \cite{attard2012non,swendsen2017thermodynamics,lebon2008understanding,lieb2003entropy,lieb1999physics,ottinger2005beyond,beyond2014,tome2012entropy,huang2008statistical}.
For weak correlations between OQS and its environment (reservoirs), SLoT can be rewritten as the Clausius inequality \cite{spohn1973irreversible,alicki2018introduction,potts2021thermodynamically} that restricts possible evolution of the system \cite{niedenzu2018quantum,bekenstein1974generalized,vedral2009entanglement,beyond2014,endres2017entropy,kammerlander2016coherence,vinjanampathy2016quantum,huang2008statistical}.
Thus, the relation between these restrictions (particularly limitations on the cooling) and microscopic quantum dynamics is of interest.

To explore that, we consider an OQS coupled to several reservoirs  $R_j$ ($j\ {\rm runs\ from\ }1{\rm\ to\ }n$) with different temperatures and chemical potentials.
The corresponding Hamiltonian reads~\cite{hofer2017markovian,malavazi2024detuning,vovchenko2024transient}
\begin{equation}
    \hat H= \hat H_S + \sum\limits_{j=1}^n\hat H_{R_j} + \sum\limits_{j=1}^n\hat H_{S R_j}.
\end{equation}
Here $\hat H_S$ is the OQS's Hamiltonian, $\hat H_{R_j}$ is the $j$-th reservoir Hamiltonian, $\hat H_{S R_j}$ describes the interaction between the OQS and $j$-th reservoir.
Markovian, linear, completely positive, and trace-preserving dynamics (MLCpTp) of the OQS's density matrix $\hat{\rho}_S$, occurring when the number of reservoirs' degrees of freedom is much larger than the one of the OQS~\cite{davies1974markovian,breuer2002theory,shishkov2019relaxation}, can be described via the Lindblad master equation (ME) \cite{gorini1976completely,potts2021thermodynamically,lindblad1976generators}
\begin{gather}\label{ME}
	\frac{{\partial \hat \rho_S}}{{\partial t}} =  - i[ {{{\hat H}_S},\hat \rho_S } ] + \sum_{j=1}^n D_j[\hat \rho_S] + \sum_{j=1}^n FS_j[\hat \rho_S].
\end{gather}
Here $D_j[\hat \rho_S]$ is a dissipator, and $FS_j[\hat \rho_S]$ is an operator that defines the OQS's frequency shifts, both produced by the interaction with the $j$-th reservoir.
A particular form of ME depends on the types of OQS-reservoirs interactions and the approach used for the dissipation description~\cite{shishkov2019relaxation,vovcenko2021dephasing,hofer2017markovian,vovchenko2021model}.

For the dissipators written in Davies form~\cite{davies1974markovian,breuer2002theory,shishkov2019relaxation}, Eq.~(\ref{ME}) implies the total density matrix to be of the form $\hat\rho (t)=\hat{\rho}_S (t) \otimes \hat{\rho}_{R_1} \otimes \cdots \otimes \hat{\rho}_{R_n}$~\cite{breuer2002theory,carmichael2009open}, where reservoirs are in Gibbs states, $\hat\rho_{R_j}=\hat\rho_{th,j}=\exp{\left(\frac{\mu_j \hat N_{R_j} - \hat H_{R_j}}{T_j}\right)}/Z_j$.
Here $T_j, \mu_j$ are $j$-th reservoir temperature and chemical potential, $Z_j=\mathrm{tr}\,e^{\frac{\mu_j \hat N_{R_j} - \hat H_{R_j}}{T_j}}$, operator $\hat N_{R_j}$ describes the number of particles in $j$-th reservoir.

Using the decrease of relative entropy in MLCpTp dynamics \cite{lindblad1975completely}, we generalize the approach from \cite{spohn1978entropy} and derive the Clausius inequality (see Supplement Materials A),
\begin{equation}\label{Clausius_EQ}
	\dot S_S + \sum_{j=1}^n \frac{\mu_j P_j - J_j}{T_j} \ge 0,
\end{equation}
for an OQS interacting with reservoirs that have different chemical potentials and temperatures.
Here $S_S=\mathrm{tr}(\hat{\rho}_S \mathrm{ln} \hat{\rho}_S)$ is OQS's entropy, $P_j=\mathrm{tr}(D_j[\hat\rho_S] \hat N _S)$ and $J_j=\mathrm{tr}(D_j[\hat\rho_S]\hat H_S)$ are particle and energy flows out of the $j$-th reservoir ($J_j>0,\ P_j>0$ means that the $j$-th reservoir loses energy and particles).
Given the form of Eq.~(\ref{Clausius_EQ}), we introduce $Y_j=-\mu_j P_j + J_j$ as generalized heat flows.
They represent total energy flow minus energy flow transmitted by the particles.
Note that flows in Eq. (\ref{Clausius_EQ}) represent reservoirs' entropies time derivatives because $\dot{S}_j=-\mathrm{tr}(\dot{\hat{\rho}}_{R_j}\mathrm{ln}\hat{\rho}_{th,j})=(\mu_j P_j - J_j)/T_j$, where $S_j=-\mathrm{tr}(\hat{\rho}_{R_j}\mathrm{ln}\hat{\rho}_{R_j})$ is $j$-th reservoir entropy. 



There are three characteristic time scales in MLCpTp dynamics: (i) $\tau_R=\max_j \tau_{R_j}$, where $\tau_{R_j}$ is the time of $j$-th reservoir occupancies redistribution due to its internal thermalization, (ii) time of OQS dissipation $\tau_S$, (iii) time of the reservoirs' temperatures and chemical potentials equalization $\tau_{eq}$ \cite{vovchenko2024transient}.

For many realizations of OQS, it holds $\tau_R\ll \tau_S \ll \tau_{eq}$~\cite{vovchenko2024transient, alicki2018introduction,haake1973statistical,saaskilahti2013thermal,hofer2017markovian,cattaneo2019local}.
This means that reservoirs' temperatures and chemical potentials are established during the shortest timescale $\tau_R$. During the time $\tau_S$, the OQS transits to a quasi-stationary state $\hat{\overline{\rho}}_S=\hat{\overline{\rho}}_S(T_1,\ldots,T_n,\mu_1,\ldots,\mu_n)$ defined by the established reservoirs' temperatures and chemical potentials~\cite{carmichael2009open,breuer2002theory,haake1973statistical,vovchenko2024transient}.
After that quasi-stationary energy, particle and heat flows out of the reservoirs are formed \cite{potts2021thermodynamically,vovcenko2023energy,vovchenko2024transient,hofer2017markovian}.
The reservoirs' temperatures and chemical potentials evolve on the $\tau_{eq}$ time scale.
At this timescale, they should be considered as time functions $\mu_j=\mu_j(t),\ T_j=T_j(t)$ (the state of the OQS is determined by the stationary solution of Eq.~(\ref{ME}) for current reservoirs' temperatures and chemical potentials).

Hence, for $t\gg \tau_S,\ \dot S_S\approx 0$, 
and Eq.~(\ref{Clausius_EQ}) simplifies to
$\sum_{j=1}^n \overline{Y}_j/T_j\le 0$.
Here $\overline{Y}_j=\overline{J}_j-\mu_j \overline{P}_j$, $\overline{P}_j=\mathrm{tr}(D_j[\hat{\overline{\rho}}_S] \hat N _S) =\overline{P}_j(T_1(t),\dots,T_n(t),\mu_1(t),\dots,\mu_n(t))$, $ \overline{J}_j=\mathrm{tr}(D_j[\hat{\overline{\rho}}_S] \hat H _S)=\overline{J}_j(T_1(t),\dots,T_n(t),\mu_1(t),\dots,\mu_n(t))$ are mentioned quasi-stationary flows.
They change energy $E_j$ and number of particles $N_j$ in the reservoirs
\begin{equation}\label{Base_eqs}
\frac{dN_j(t)}{dt}=-\overline{P}_j(t),\ \ \ \frac{dE_j(t)}{dt}=-\overline{J}_j(t).
\end{equation}
Here, $N_j(t)=N_j(T_j(t),\mu_j(t)),\ E_j(t)=E_j(T_j(t),\mu_j(t))$ are known functions of chemical potentials and temperatures \cite{Landau_StatPhys,zaitsev2008introduction}.
Thus, we can rewrite the system~(\ref{Base_eqs}) as 
\begin{equation}\label{Base_eqs_Tmu}
    \begin{pmatrix}
        \dot{T}_j \\
        \dot{\mu}_j
    \end{pmatrix}=-
    \begin{pmatrix}
        \partial N_j/\partial T_j & \partial N_j/\partial \mu_j\\
        \partial E_j/\partial T_j & \partial E_j/\partial \mu_j
    \end{pmatrix}^{-1}
    \begin{pmatrix}
        \overline{P}_j \\
        \overline{J}_j
    \end{pmatrix},
\end{equation}
if the Jacobian of Eqs.~(\ref{Base_eqs}) does not equal zero.
Note that total energy and number of particles in reservoirs are preserved in Eqs.~(\ref{Base_eqs})-(\ref{Base_eqs_Tmu}): $\sum_j \overline{J}_j(t)=0,\,\sum_j \overline{P}_j(t)=0$ as the OQS is in the stationary state of Eq~(\ref{ME}).
Hence, $\sum_j N_j(t)=\mathrm{const},\ \sum_j E_j(t)=\mathrm{const}$.
Thus, if a thermodynamical equilibrium state of Eqs. (\ref{Base_eqs})-(\ref{Base_eqs_Tmu}) exists, it satisfies
\begin{gather}\label{Eq_T_mu}
    \sum_j N_j(T_{eq},\mu_{eq})=\sum_j N_j(T_j(0),\mu_j(0)), \\ \nonumber
    \sum_j E_j(T_{eq},\mu_{eq})=\sum_j E_j(T_j(0),\mu_j(0)).
\end{gather}
Here $T_{eq},\,\mu_{eq}$ are equilibrium temperature and chemical potential.
Note that the equilibrium state of Eqs. (\ref{Base_eqs_Tmu}) also implies $\overline{P}_j=0,\,\overline{J}_j=0$.


We should highlight the restrictions imposed on Eqs.~(\ref{Base_eqs_Tmu}) by SLoT.
For reservoirs with equal chemical potentials, we have $\sum_j \overline{Y}_j=0$.
Let number the reservoirs in increasing order of their temperatures: $T_1\le \cdots \le T_n$.
For two reservoirs, $\overline{Y}_1/T_1 + \overline{Y}_2/T_2\le 0,\ \overline{Y}_1+\overline{Y}_2=0$, and we get the well-known result $\overline{Y}_2\ge 0$, i.e., the heat flows from the hot reservoir to the cold one at all moments of time~\cite{adkins1983equilibrium,kondepudi2014modern,lebon2008understanding}.
For more reservoirs, from $\sum_{j=1}^n \overline{Y}_j/T_j\le 0$ and $\sum_j \overline{Y}_j=0$ we get
\begin{gather}
    \sum_{j=2}^n \overline{Y}_j \left(\frac{1}{T_j}-\frac{1}{T_1}\right)\le 0,\ \ 
    \sum_{j=1}^{n-1} \overline{Y}_j \left(\frac{1}{T_j}-\frac{1}{T_n}\right)\le 0.
\end{gather}
Hence, there exists $\overline{Y}_j\ge 0$ with $T_j>T_1$, and there exists $\overline{Y}_j\le 0$ with $T_j<T_n$.
Thus, for reservoirs with equal chemical potentials, from SLoT in the stationary regime follows: at least one of the not the coldest reservoirs loses heat, and at least one of the not the hottest reservoirs receives heat.
Hence, SLoT does not prohibit cooling (heating) of the coldest (hottest) reservoir without external influence on the whole system.
For instance, if $\overline{Y}_1=1,\,\overline{Y}_2=-5$, $\overline{Y}_3=20,\,\overline{Y}_4=-16$,\,$T_1=10,\,T_2=20,\,T_3=60,\,T_4=70$ (dimensionless units) are realized, one has $\sum_1^4\overline{Y}_j=0,\,\sum_1^4\overline{Y}_j/T_j\le 0$, and the coldest reservoir loses heat while the hottest one receives it.

If chemical potentials of reservoirs are not equal, we have $\sum_j \overline{Y}_j\ne 0$ in the general case.
Hence, there is a hyperplane in space $\{\overline{P}_2,\ldots,\overline{P}_n,\overline{J}_2,\ldots,\overline{J}_n\}$ that separates allowed transport regimes from forbidden ones:
$\sum\limits_{j=2}^n (\mu_j/T_j-\mu_1/T_1)\overline{P}_j-\sum\limits_{j=2}^n (1/T_j-1/T_1)\overline{J}_j=0$.


Further, we explicitly demonstrate the possibility of the coldest reservoir cooling.
Let us consider a quadratic fermionic Hamiltonian ($\hbar=1$)
\cite{yang2013master,landovitz1967quadratic,nava2023lindblad1,zaitsev2014introduction,zimbovskaya2011electron,zatsarynna2024many,kohler2005driven,haug2008quantum}
\begin{equation}\label{HS}
    \hat H_S=\sum_{v,w} \varepsilon_{vw} \hat c^\dag_v \hat c_w=\sum_\kappa \omega_\kappa \hat a^\dag_\kappa \hat a_\kappa.
\end{equation}
Here $\varepsilon_{vw}=\varepsilon_{wv},\ \hat c_v$ are annihilation operators, and $\hat a_\kappa$ are annihilation operators that diagonalize $\hat{H}_S$, $\{\hat c_v,\hat c_w\}=\{\hat a_v,\hat a_w\}=0$, $\{\hat c^\dag_v,\hat c_w\}=\{\hat a^\dag_v,\hat a_w\}=\delta_{vw}$. 
ME in Davies form for this system coupled to several reservoirs of dissipative type reads \cite{potts2021thermodynamically,barthel2022solving,nava2023lindblad2,nava2024dissipation}:
\begin{gather}\label{ME_ferm}
    \frac{{\partial \hat \rho_S}}{{\partial t}} =  - i[ {{{\hat H}_S},\hat \rho_S } ] + \sum_{j=1}^n \sum_\kappa \Lambda_{\kappa,j}[\hat \rho_S],\\ \nonumber
	\Lambda_{\kappa,j}[\hat \rho_S] =\\ \nonumber
 =\frac{{{G_j}( - {\omega_\kappa }, \mu_j, T_j)}}{2}\hat L[\hat a_\kappa,\hat a^\dag_\kappa] +\frac{{{G_j}({\omega_\kappa}, \mu_j, T_j)}}{2}\hat L[\hat a^\dag_\kappa,\hat a_\kappa].
\end{gather}
Here $\hat L [\hat X, \hat Y] = 2\hat X \hat \rho_S \hat Y-\hat Y \hat X \hat\rho_S -\hat\rho_S \hat Y \hat X$ is a Lindblad superoperator, ${G_j}(\omega,\mu_j, T_j)=\gamma_j(\omega)n_j(\omega,\mu_j,T_j)$ are reservoirs' correlation functions, $n_j(\omega,\mu_j,T_j)=1/(e^ {(\omega-\mu_j)/T_j}+1)$, $\gamma_j\left(\omega\right) = \pi g_j\left(\omega\right) |\chi_{\omega,j}\left(\omega\right)|^2$, $g_j\left(\omega\right)$ are $j$-th reservoir's occupancy, dissipation rates, and density of states at frequency $\omega$, respectively, $\chi_{\omega,j}\left(\omega\right)$ are interaction constants between the OQS and $j$-th reservoir, ${G_j}(-\omega,\mu_j,T_j)=\gamma_j(\omega)(1-n_j(\omega,\mu_j,T_j))$, $\omega>0$, $D_j[\hat \rho_S]=\sum_\kappa \Lambda_{\kappa,j}[\hat \rho_S]$.

From Eq.~(\ref{ME_ferm}), equations on the dynamics of $\langle \hat{a}^\dag_\kappa \hat{a}_\kappa \rangle$ and their stationary solutions can be found.
Substituting them in $\langle \dot{H}_S\rangle$, we get stationary energy, particles, and heat flows from reservoirs
\begin{gather}\label{J_stat}
    \overline{J}_j=\sum_\kappa \omega_\kappa \gamma_j (\omega_\kappa)(n_j(\omega_\kappa,\mu_j, T_j) - \tilde{n}(\omega_\kappa)),\\ \nonumber
    \overline{P}_j=\sum_\kappa \gamma_j (\omega_\kappa)(n_j(\omega_\kappa,\mu_j, T_j) - \tilde{n}(\omega_\kappa)), \\ \nonumber
    \overline{Y}_j=\sum_\kappa (\omega_\kappa - \mu_j)\gamma_j (\omega_\kappa)(n_j(\omega_\kappa,\mu_j, T_j) - \tilde{n}(\omega_\kappa)).
\end{gather}
Here $\tilde{n}(\omega_\kappa)=\sum_{j=1}^n \gamma_j(\omega_\kappa)n_j(\omega_\kappa,\mu_j, T_j)/\sum_{j=1}^n \gamma_j(\omega_\kappa)$.
Thus, $\tilde{n}(\omega_\kappa)$ can be interpreted as weighted occupancy at frequency $\omega_\kappa$~\cite{vovchenko2024transient}.
In Supplement Material C we show that Eqs.~(\ref{J_stat}) satisfy SLoT.

\begin{table}
\caption{Energy, particle, and heat flows (Eqs.~(\ref{J_stat})) signs' dependencies on the frequency $\omega$ (see also \cite{potts2021thermodynamically,potts2024quantum,topp2015steady,jiang2015phonon}).}
\label{Tabs}
\begin{minipage}{\linewidth}
\centering
\begin{tabular}{|c|c|c|c|c|}
    \hline
    $T_1<T_2, \mu_1>\mu_2$ & $\omega<\mu_2$ & $\mu_2<\omega<\mu_1$ & $\,\mu_1<\omega<\tilde{\omega}\,$ & $\,\tilde{\omega}<\omega\,$\\ 
    \hline
    $\overline{j}_1 (\omega)$ & \,+\, & \,+\,  & \,+\,  & \,--\, \\
    \hline
    $\overline{p}_1 (\omega)$ & + & +  & +  & -- \\
    \hline
    $\overline{y}_1 (\omega)$ & -- & --  & +  & -- \\
    \hline
    $\overline{j}_2 (\omega)$ & -- & --  & -- & +\\
    \hline
    $\overline{p}_2 (\omega)$ & -- & --  & --  & +\\
    \hline
    $\overline{y}_2 (\omega)$ & + & --  & --  & +\\
     \hline
\end{tabular}
\end{minipage}
\vspace{0.1cm}

\begin{minipage}{\linewidth}
\centering
\begin{tabular}{|c|c|c|c|c|}
    \hline
    $T_1>T_2, \mu_1>\mu_2$ & $\,\omega<\tilde{\omega}\,$& $\,\tilde{\omega}<\omega<\mu_2\,$ & $\mu_2<\omega<\mu_1$ & $\mu_1<\omega$ \\
    \hline
    $\overline{j}_1 (\omega)$ & \,--\, & \,+\,  & \,+\,  & \,+\, \\
    \hline
    $\overline{p}_1 (\omega)$ & -- & +  & +  & + \\
    \hline
    $\overline{y}_1 (\omega)$ & + & --  & --  & + \\
    \hline
    $\overline{j}_2 (\omega)$ & + & --  & --  & --\\
    \hline
    $\overline{p}_2 (\omega)$ & + & --  & --  & --\\
    \hline
    $\overline{y}_2 (\omega)$ & -- & +  & --  & --\\
     \hline
\end{tabular}
\end{minipage}
\end{table}

Substituting Eqs.~(\ref{J_stat}) in Eqs.~(\ref{Base_eqs_Tmu}), we get (see Supplement Materials B)
\begin{gather}\label{T_dyn}
\dot T_j = - \frac{-Q_{\alpha_j} T_j \overline{P}_j + Q_{\alpha_j-1}  \overline{J}_j }{ \tilde{A}_j V_j T_j^{\alpha_j} x_j^{2\alpha_j-4}\left(x_j^2-\cfrac{\pi^2}{6}\alpha_j(\alpha_j-1)\right)},\\ \nonumber
\dot x_j =  \frac{-Q_{\alpha_j+1} T_j \overline{P}_j +Q_{\alpha_j}  \overline{J}_j }{ \tilde{A}_j V_j T_j^{\alpha_j+1} x_j^{2\alpha_j-4}\left(x_j^2-\cfrac{\pi^2}{6}\alpha_j(\alpha_j-1)\right)}.
\end{gather}
Here $x_j=\mu_j/T_j,\,Q_{\alpha_j}=x_j^{\alpha_j} + \cfrac{\pi^2}{6}\alpha_j (\alpha_j -1) x_j^{\alpha_j-2},\,\tilde{A}_j$ are constants, $\alpha_j=D_j/d_j,\,D_j$ is $j$-th reservoir dimension with energy spectrum $\varepsilon=c p^{d_j}$.
Eqs.~(\ref{T_dyn}) have a unique equilibrium state at which all temperatures and chemical potentials are equal $T_{eq}$ and $\mu_{eq}$, respectively (see Supplement Materials C, D and Figs.~\ref{2Res_1}-\ref{4Res_1}) that is confirmed by numerical simulations.

Flows' amplitudes are proportional to
$ n_j(\omega_\kappa,\mu_j, T_j)-\tilde{n}(\omega_\kappa)$.
Hence, the reservoir that is the most (least) populated at the frequency $\omega_\kappa$ loses (receives) both energy and particles at this frequency.
Thus, terms in the sums in Eqs.~(\ref{J_stat}) can have different signs.

In the case of two reservoirs (by definition of $\tilde{n}(\omega_\kappa)$), expression $\gamma_1(\omega_\kappa)( n_1(\omega_\kappa,\mu_1, T_1)-\tilde{n}(\omega_\kappa))$ equals
$\Gamma_{\omega_\kappa}(n_1(\omega_\kappa,\mu_1, T_1) - n_2(\omega_\kappa,\mu_2, T_2))\equiv \overline{p}_{1}(\omega_\kappa)$,
where $\Gamma_{\omega_\kappa}=\gamma_1 (\omega_\kappa)\gamma_2(\omega_\kappa)/(\gamma_1 (\omega_\kappa)+\gamma_2(\omega_\kappa))$.
Occupancies of reservoirs can be equal at some frequency $\tilde{\omega}$ determined by $\tilde{\omega}=(T_1 \mu_2 - T_2 \mu_1)/(T_1-T_2)$.
Let $\mu_1>\mu_2$.
A positive $\tilde{\omega}$ exists, if $(\mu_2-(T_2/T_1)\mu_1)(1-T_2/T_1)>0$ (note that either $\tilde\omega > \mu_1$ or $\tilde\omega < \mu_2$).
This condition is always fulfilled, if $T_2>T_1$, and it is broken, if $T_2/T_1<1,\ (T_2/T_1)(\mu_1/\mu_2)>1$.
In the latter case, the occupancies are different at all frequencies, hence, energy and particles transit from the hot reservoir to the cold one at all frequencies.

\begin{figure}
\begin{minipage}[h]{\linewidth}
\center{\includegraphics[width=0.9\linewidth]{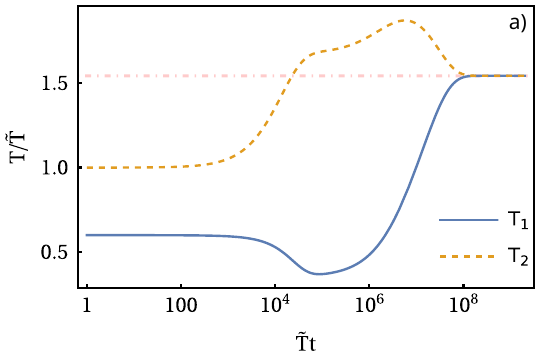}}
\end{minipage}
\begin{minipage}[h]{\linewidth}
\center{\includegraphics[width=0.9\linewidth]{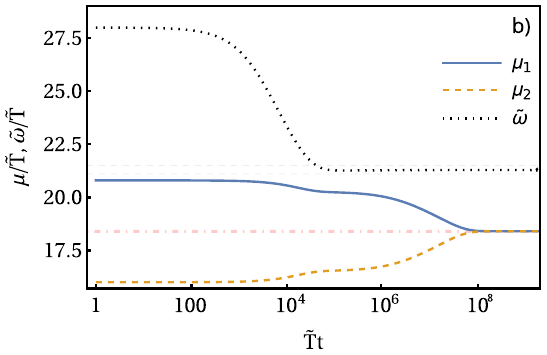}}
\end{minipage}
\caption{Dynamics of a) temperatures and b) chemical potentials of two reservoirs coupled to OQS with states $\omega_1/\mathrm{\tilde{T}}=21.1,\,\omega_2/\mathrm{\tilde{T}}=21.5$ depicted by dashed horizontal lines in b) ($\mathrm{\tilde{T}}=1000$~K).
Initial temperatures and chemical potentials are $T_1/\mathrm{\tilde{T}}=0.6,\,T_2/\mathrm{\tilde{T}}=1.0,\,\mu_1/\mathrm{\tilde{T}}=20.8,\,\mu_2/\mathrm{\tilde{T}}=16.0,\,\gamma_{j}(\omega)=10^{-4}(\omega/\mathrm{\tilde{T}})^{\alpha_{j}-1}$, $\alpha_{j}=3/2$.
Dot-dashed horizontal lines are equilibrium a) temperature and b) chemical potential from Eqs.~(\ref{Eq_T_mu}).
Dotted line depicts $\tilde{\omega}$ dynamics.}
\label{2Res_1}
\end{figure}

Let introduce
$\overline{p}_{2}(\omega_\kappa)=-\overline{p}_{1}(\omega_\kappa),\ \overline{j}_{1,2}(\omega)=\omega \overline{p}_{1,2}(\omega)$,
$\overline{y}_{1,2}(\omega)=(\omega -\mu_{1,2})\overline{p}_{1,2}(\omega)$.
These functions reproduce frequency dependencies of energy, particle, and heat flows' signs, see Tab.~\ref{Tabs}.
Functions $\overline{p}_{1,2}(\omega),\ \overline{j}_{1,2}(\omega)$ change signs at $\omega=\tilde{\omega}$.
Functions $\overline{y}_{1,2}(\omega)$ change signs at $\omega=\mu_{1,2}$ and $\omega=\tilde{\omega}$.
Further, we discuss the most interesting regimes from Tab.~\ref{Tabs}.

\begin{figure}
\begin{minipage}[h]{\linewidth}
\center{\includegraphics[width=0.9\linewidth]{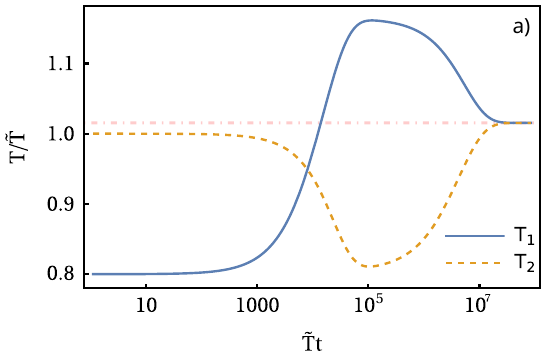}}
\end{minipage}
\begin{minipage}[h]{\linewidth}
\center{\includegraphics[width=0.9\linewidth]{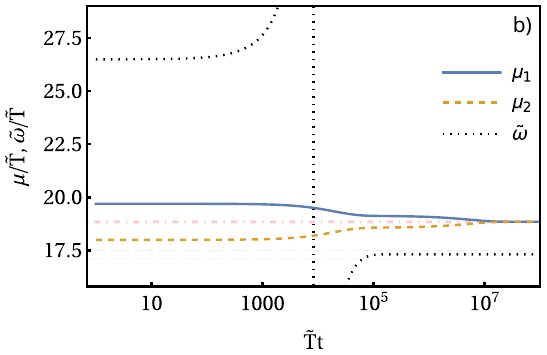}}
\end{minipage}
\caption{Same as in Fig. \ref{2Res_1} but for $\omega_1/\mathrm{\tilde{T}}=17.1,\,\omega_2/\mathrm{\tilde{T}}=17.5$, $T_1/\mathrm{\tilde{T}}=0.8,\,T_2/\mathrm{\tilde{T}}=1.0,\,\mu_1/\mathrm{\tilde{T}}=18.8,\,\mu_2/\mathrm{\tilde{T}}=18.0$.}
\label{2Res_2}
\end{figure}

As reservoirs' temperatures and chemical potentials depend on time $\tilde{\omega}=\tilde{\omega}(t)$.
Suppose that initially $T_1<T_2,\ \mu_2<\mu_1$, and all $\omega_\kappa$ satisfy $\mu_1<\omega_\kappa<\tilde{\omega}$.
Then, the cold reservoir loses energy, particles, and heat while the hot one gains them (see Tab.~\ref{Tabs}).
For large $x_j$ in Eqs.~(\ref{T_dyn}) we get $\dot{T}_j \approx - \overline{Y}_j/\tilde{A}_j V_j T_j^{\alpha_j} x_j^{\alpha_j - 1}$.
Thus, we can expect the coldest reservoir cooling.
Numerical simulations confirm this fact, see Fig.~\ref{2Res_1}a.
The coldest reservoir is cooling down initially: $\min_t T_{cold} (t)/ T_{cold}(0)\approx 0.6$ (here $T_{cold}$ is the temperature of the coldest reservoir).

However, as noted above, there exists the equilibrium state of Eqs.~(\ref{Base_eqs_Tmu}) with equal temperatures and chemical potentials.
It can not be reached with only these flows.
Hence, the evolution of temperatures and chemical potentials should dynamically modify the flows to those that lead to the equalization of reservoirs' temperatures and chemical potentials.
This happens as follows.

After a certain moment of time, the $\tilde{\omega}$ starts satisfying the $\omega_1<\tilde{\omega}<\omega_2$ condition (see Fig.~\ref{2Res_1}b).
Further, the system supports a mixed transport regime, i.e., a combination of transport regimes from Tab.~\ref{Tabs}.
In this regime, terms under the sums in Eqs.~(\ref{J_stat}) have opposite signs.
This diminishes total energy and particle flows and leads the total system to the mentioned equilibrium state despite the previous non-trivial dynamics.

Values of equilibrium temperature and chemical potential coincide with those calculated via Eqs.~(\ref{Eq_T_mu}).
Notably, equilibrium temperature in Fig.~\ref{2Res_1}a is higher than all initial reservoirs' temperatures.
This happens due to the equalization of chemical potentials.
Thus, it is possible to temporarily cool down the coldest reservoir with a subsequent increase in its temperature up to the equilibrium temperature that exceeds the initial one.

\begin{figure}
\begin{minipage}[h]{\linewidth}
\center{\includegraphics[width=0.9\linewidth]{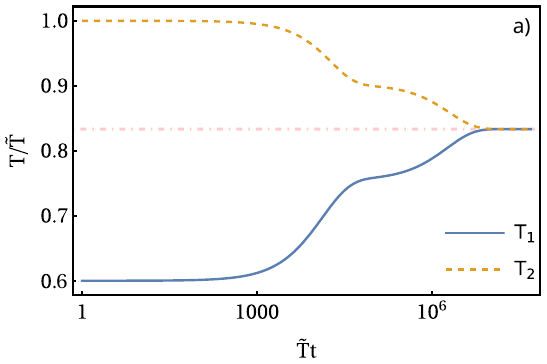}}
\end{minipage}
\begin{minipage}[h]{\linewidth}
\center{\includegraphics[width=0.9\linewidth]{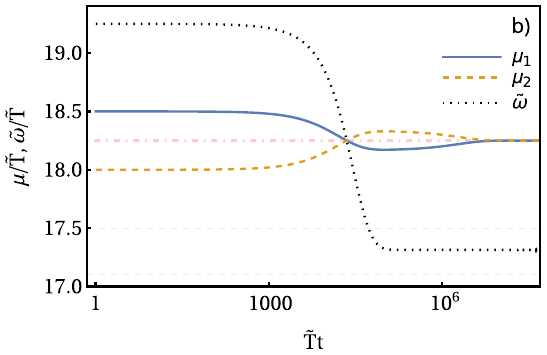}}
\end{minipage}
\caption{Same as in Fig. \ref{2Res_1} but for $\omega_1/\mathrm{\tilde{T}}=17.1,\,\omega_2/\mathrm{\tilde{T}}=17.5$, $T_1/\mathrm{\tilde{T}}=0.8,\,T_2/\mathrm{\tilde{T}}=1.0,\,\mu_1/\mathrm{\tilde{T}}=18.8,\,\mu_2/\mathrm{\tilde{T}}=18.0$.}
\label{2Res_3}
\end{figure}

If initially $T_1<T_2,\ \omega_\kappa<\mu_2<\mu_1<\tilde{\omega}$, the reservoirs can dynamically change their orderliness in temperatures (Fig.~\ref{2Res_2}a).
This leads to a singularity in $\tilde{\omega}$: $\tilde{\omega}$ goes through $+\infty$ to $-\infty$ (Fig.~\ref{2Res_2}b).
Hence, there is a time interval during which $\tilde{\omega}$ is negative.
Thus, within this time interval, reservoirs' occupancies are different at all frequencies.
In addition, a similar regime can change the reservoirs' orderliness in chemical potentials (Fig.~\ref{2Res_3}b).
When chemical potentials' coincide, we get $\mu_1=\mu_2=\tilde{\omega}$.
Thus, in both cases $\tilde{\omega}$ transits from $\tilde{\omega}>\mu_1$ to $\tilde{\omega}<\mu_2$ bypassing the forbidden range between $\mu_2$ and $\mu_1$ to finally satisfy the mentioned condition $\omega_1<\tilde{\omega}<\omega_2$.
After that, a mixed transport regime is formed, and the total system reaches the equilibrium state defined by Eqs.~(\ref{Eq_T_mu}).

In the discussed regimes, temperatures and chemical potentials exhibit non-monotonic behavior.
This indicates transformations of transport regimes due to the evolution of reservoirs' temperatures and chemical potentials.
For more reservoirs, this non-monotonic behavior becomes intricate.
In Fig.~\ref{4Res_1}, it is represented for a case of four reservoirs.
As previously, temperatures and chemical potentials exhibit non-monotonic behavior but have several local extrema.
Finally, the system reaches equilibrium state.
Equilibrium temperature and chemical potential are defined by Eqs.~(\ref{Eq_T_mu}).
The equilibrium temperature here is also higher than all initial reservoirs' temperatures (see Supplement Materials B for estimations of equilibrium temperature and chemical potential).

\begin{figure}
\begin{minipage}[h]{\linewidth}
\center{\includegraphics[width=0.9\linewidth]{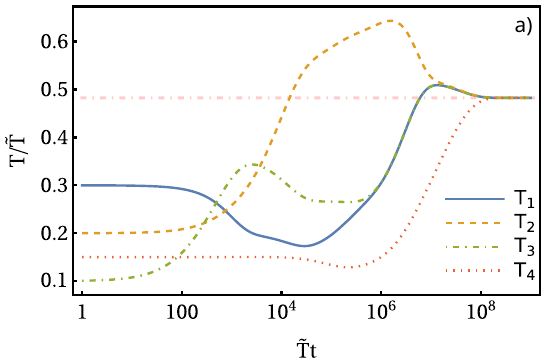}}
\end{minipage}
\begin{minipage}[h]{\linewidth}
\center{\includegraphics[width=0.92\linewidth]{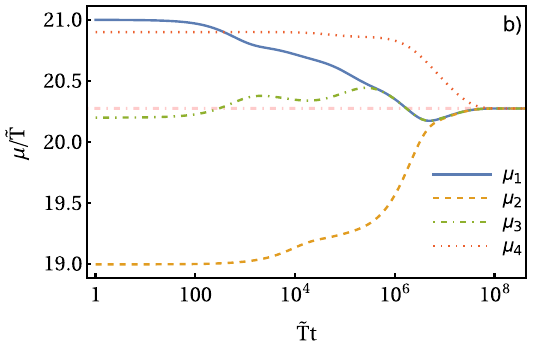}}
\end{minipage}
\caption{Same as in Fig. \ref{2Res_1} but for the case of four reservoirs: $\omega_1/\mathrm{\tilde{T}}=21.5,\,\omega_2/\mathrm{\tilde{T}}=21.2,\,T_1/\mathrm{\tilde{T}}=0.3,\,T_2/\mathrm{\tilde{T}}=0.2$, $T_3/\mathrm{\tilde{T}}=0.1,\,T_4/\mathrm{\tilde{T}}=0.15,\,\mu_1/\mathrm{\tilde{T}}=21.0,\,\mu_2/\mathrm{\tilde{T}}=19.0,\,\mu_3/\mathrm{\tilde{T}}=20.2$, $\mu_4/\mathrm{\tilde{T}}=20.9$, $\gamma_{1}(\omega)=\gamma_{3}(\omega)=10\gamma_{2}(\omega),\,\gamma_{4}(\omega)=0.05\gamma_{2}(\omega)$.}
\label{4Res_1}
\end{figure}


To sum up, we derived the SLoT in the form of the Clausius inequality for an OQS connected to several reservoirs with different temperatures and chemical potentials, supposing MLCpTp dynamics of OQS.
We showed that SLoT in the form of the Clausius inequality does not prohibit heat transport from the coldest reservoir to the hottest one in the absence of external refrigeration.
We have provided an example of such a system.
We showed that in this system the coldest reservoir can be cooled down locally in time, reservoirs' temperatures and chemical potentials exhibit non-monotone behavior, and the final equilibrium temperature can be higher than all initial reservoirs' temperatures.
Also, we presented non-trivial regimes in which reservoirs' change their orderliness in temperatures and chemical potentials.
Achieved results can be used in the design of
quantum thermal machines~\cite{binder2018thermodynamics,bhattacharjee2021quantum,kosloff2014quantum} and nanoelectronic devices~\cite{gordeeva2020record,islam2024evaluating}.
\\

\textbf{Acknowledgement.}
This work was financially supported by the Russian Science Foundation (project no. 20-72-10057).
I.V.V. and E.S.A. acknowledge the support of the Foundation for the Advancement of Theoretical Physics and Mathematics BASIS.

\bibliography{Thermodynamics_Topics}

\section*{Supplement materials}

\subsection{Derivation of the Clausius inequality for the case of reservoirs with different chemical potentials}\label{SpL_A}
In this section, we consider an open quantum system (OQS) connected to several reservoirs with different temperatures and chemical potentials.
We derive the second law of thermodynamics (SLoT) in Clausius form for this system considering its Markovian, linear, completely positive, and trace-preserving dynamics (MLCpTp).
The presented proof is a natural generalization of the one from \cite{spohn1978entropy}.
It accounts for the case of reservoirs with different chemical potentials.
We show that SLoT can be seen as a consequence of the zero law of thermodynamics and MLCpTp dynamics.

In the main text, the Hamiltonian of the OQS equals
\begin{equation}
    \hat H= \hat H_S + \sum\limits_{j=1}^n\hat H_{R_j} + \sum\limits_{j=1}^n\hat H_{S R_j}.
\end{equation}

Irreversible MLCpTp dynamics of the OQS connected with $n$ reservoirs can be described by the Lindblad master equation (ME)~\cite{breuer2002theory,shishkov2019relaxation,gorini1976completely,davies1974markovian}
\begin{gather}\label{ME_SpL}
	\frac{{\partial \hat \rho_S}}{{\partial t}} =  - i[ {{{\hat H}_S},\hat \rho_S } ] + \sum_{j=1}^n D_j[\hat \rho_S] + \sum_{j=1}^n FS_j[\hat \rho_S].
\end{gather}

In the case of one reservoir (with index $j$), a thermodynamically compatible ME should lead to the OQS evolving to the Gibbs state $\hat\rho_{S,j}=\exp{\left(\frac{\mu_j \hat N_{S} - \hat H_{S}}{T_j}\right)}/\mathrm{tr}\left(\exp{\left(\frac{\mu_j \hat N_{S} - \hat H_{S}}{T_j}\right)}\right)$, $\hat N_{S}$ is operator that describes the number of particles in the OQS \cite{Landau_StatPhys,zaitsev2008introduction,deffner2019quantum,spohn1978entropy,huang2008statistical}.
Hence, for every $j$, it holds
\begin{equation}
    0= - i[ {{{\hat H}_S},\hat\rho_{S,j}} ] +  D_j[\hat\rho_{S,j}] +  FS_j[\hat\rho_{S,j}].
\end{equation}

We assume that the internal dynamic of the OQS does not change the number of particles, i.e., $[\hat N_S,\hat H_S]=0$.
Hence, $[ {{{\hat H}_S},\hat\rho_{S,j}} ] =0$.
Also, as $FS_j[\cdot]$ are responsible for the frequency shifts, they can be added to the Hamiltonian $\hat H_S$ to form an effective Hamiltonian $\hat H_{eff}$ that has the same eigenstates as the Hamiltonian $\hat H_S$.
Hence, $FS_j[\hat\rho_{S,j}]=0$.
Thus, in a thermodynamically compatible ME, $D_j[\hat\rho_{S,j}]=0$.

From $\mathrm{tr}([\hat\rho_S,\hat H_{eff}])=0$ it follows that for the dynamics to be trace-preserving, it is needed that $\mathrm{tr}(D_j[\hat\rho_S])=0$.

Eq.~(\ref{ME_SpL}) can be rewritten as $\partial \hat \rho(t)/\partial t = \Lambda [\hat \rho (t)$]. Here $\Lambda[\cdot]$ is a superoperator that gives the right-hand side of the Eq.~(\ref{ME_SpL}), being applied to the density matrix.
As $\Lambda$ does not depend on time, the solution of this equation is
$\hat\rho (t)=\exp{((t-t_0)\Lambda)}\hat\rho(t_0)$.
This can be rewritten as $\hat\rho (t)=\Phi_{t,t_0}[\hat\rho(t_0)]$, where $\Phi_{t,t_0}[\cdot]=\exp{((t-t_0)\Lambda)}[\cdot]$ is the mapping that maps the density matrix from time $t_0$ to time $t$.
This mapping is MLCpTp by the construction~\cite{breuer2002theory,shishkov2019relaxation,gorini1976completely,davies1974markovian}.
In the same manner, we can introduce mappings produced by individual reservoirs' superoperators $\Phi_{t,t_0}^{(j)}[\cdot]=\exp{((t-t_0)\Lambda_j)}[\cdot]$, $\Lambda_j[\cdot]=D_j[\cdot]+FS_j[\cdot]$ that are also MLCpTp.

We study the entropy of the OQS $S = - {\rm{tr}} \left(\hat \rho_S \ln \hat \rho_S \right)$ by considering relative entropy:
\begin{equation}\label{Sps_SpL}
	S(\hat \rho|\hat \sigma)=\mathrm{tr}(\hat \rho \mathrm{ln} \hat \rho) - \mathrm{tr}(\hat \rho \mathrm{ln} \hat \sigma).
\end{equation}
Here $\hat \rho$ and $\hat \sigma$ are density matrices of the same dimensions.
Relative entropy decreases under MLCpTp mapping being applied to both $\hat \rho$ and $\hat \sigma$, i.e., $S(O[\hat \rho]|O[\hat \sigma])\le S(\hat \rho|\hat \sigma)$, where $O[\cdot]$ is a linear, completely positive, and trace-preserving mapping \cite{lindblad1975completely}.

To describe the dynamics of the entropy in our case, it is convenient to use $\hat \rho=\hat \rho_S$, $\hat \sigma=\hat\rho_{S,j}$, and $O[\cdot]=\Phi_{t,t_0}^{(j)}[\cdot]$.
As $\Phi_{t,t_0}^{(j)}[\hat\rho_{S,j}]=\hat\rho_{S,j}$ we have
\begin{gather}
    f^{(j)}(t)\equiv S(\Phi_{t,t_0}^{(j)}[\hat\rho_S(t_0)]|\Phi_{t,t_0}^{(j)}[\hat\sigma(t_0)])=\\ \nonumber
    =S(\Phi_{t,t_0}^{(j)}[\hat\rho_S(t_0)]|\hat\rho_{S,j}).
\end{gather}

The functions $f^{\left(j\right)} \left(t\right)$ are decreasing.
Indeed
\begin{gather}
    S(\Phi_{t_2,t_0}^{(j)}[\hat\rho_S(t_0)]|\hat\rho_{S,j})=\\ \nonumber
    =S(\Phi_{t_2,t_0}^{(j)}[\hat\rho_S(t_0)]|\Phi_{t_2,t_0}^{(j)}[\hat\rho_{S,j}])=\\ \nonumber
    =S( \Phi_{t_2,t_1}^{(j)}[\Phi_{t_1,t_0}^{(j)}[\hat\rho_S(t_0)]]|\Phi_{t_2,t_1}^{(j)}[\Phi_{t_1,t_0}^{(j)}[\hat\rho_{S,j}]])\le\\ \nonumber
    \le S(\Phi_{t_1,t_0}^{(j)}[\hat\rho_S(t_0)]|\Phi_{t_1,t_0}^{(j)}[\hat\rho_{S,j}])=
    S(\Phi_{t_1,t_0}^{(j)}[\hat\rho_S(t_0)]|\hat\rho_{S,j}).
\end{gather}
Here we have used the Markovian property of $\Phi_{t,t_0}^{(j)}[\cdot]$.
Thus, $f^{(j)}(t)$ are decreasing functions, and they tend to $S\left( \hat \rho_{S,j} \left| \right. \hat \rho_{S,j} \right) = 0$, while $t \rightarrow +\infty$.
Hence, $f^{(j)}(t) \ge 0$ and $df^{(j)}/dt\le0$.
Thus,
\begin{gather}\label{ClInEq}
    \sum_j \frac{df^{(j)}}{dt}=\\ \nonumber
    =\sum_j\mathrm{tr}(\Lambda_{j}[\hat\rho_S(t)]\mathrm{ln}\hat\rho_{S})-
    \sum_j\mathrm{tr}(\Lambda_{j}[\hat\rho_S(t)]\mathrm{ln}\hat\rho_{S,j})=\\ \nonumber
    =\mathrm{tr}\bigg(\bigg(i[\hat\rho_S,\hat H_S]+\sum_j\Lambda_{j}[\hat\rho_S(t)]\bigg)\mathrm{ln}\hat\rho_{S}\bigg)-\\ \nonumber
    -\sum_j\mathrm{tr}(\Lambda_{j}[\hat\rho_S(t)]\mathrm{ln}\hat\rho_{S,j})= \\ \nonumber
    =\mathrm{tr}(\dot{\hat{\rho}}_S \mathrm{ln} \hat \rho_S)-\sum_j\mathrm{tr}(D_j[\hat\rho_S](\mu_j \hat N _S - \hat H_S)/T_j))=\\ \nonumber
    =-\dot S_S-\sum_j\frac{\mu_j P_j -J_j}{T_j}\le 0
\end{gather}

This is inequality (3) from the main text.
This form of the Clausius inequality appears in classical thermodynamics too \cite{lebon2008understanding}, where dynamics of thermodynamical values implies entropy additivity and SLoT by default, i.e., $\dot{S}_S+\sum_j \dot{S}_{j} \ge 0$.
This follows from the fact that terms under the sum in the achieved Clausius inequality~(\ref{ClInEq}) just represent $\dot{S}_j=-\mathrm{tr}(\dot{\hat{\rho}}_{R_j}\mathrm{ln}\hat{\rho}_{th,j})=(\mu_j P_j - J_j)/T_j$ as it is shown in the main text.

\subsection{The equations describing the dynamics of reservoirs' temperatures and chemical potentials}
\label{SpL_T_mu}

To establish the laws governing temperatures' and chemical potentials' dynamics, we need to find energy and the number of particles in each of the reservoirs.
For that, we use an expansion  
\begin{equation}\label{Appr_int_SpL}
\int\limits_0^{+\infty}\frac{f(\varepsilon)d\varepsilon}{\exp((\varepsilon-\mu)/T)+1}\approx \int\limits_0^{\mu}f(\varepsilon)+\frac{\pi^2}{6}T^2f'(\mu),
\end{equation}
that is valid for arbitrary function $f(\varepsilon)$, if $\mu/T\gg1$ \cite{Landau_StatPhys}.

Using this formula, we can estimate energy $E_j$ and the number of particles $N_j$ in reservoirs of ideal Fermi gas.
For a $D$-dimentional reservoir that has an energy spectrum $\varepsilon=c p^d$, the density of states equals $g(\varepsilon)=S_{D}\varepsilon^{D/d-1}/((2\pi\hbar)^D d c^{D/d})$.
Here, value
$S_{D}=D\pi^{D/2}/\Gamma(D/2+1)$ is equal to the surface area of a $D$-dimensional unit sphere.
Thus,
\begin{gather}\label{NNN_SpL}
N_j=\frac{V_j S_{D-1}}{(2\pi \hbar)^D}\int_0^{+\infty} \frac{\varepsilon^{\alpha_j-1}d\varepsilon}{\exp((\varepsilon-\mu_j)/T_j)+1}=\\ \nonumber
=A_j V_j \left(\int_0^{\mu_j} \varepsilon^{\alpha_j-1} d\varepsilon + \frac{\pi^2}{6}T^2_j (\varepsilon^{\alpha_j-1})'|_{\mu_j}\right) =\\ \nonumber
=\frac{A_j V_j}{\alpha_j}\left(\frac{\mu_j}{T_j}\right)^{\alpha_j} T_j^{\alpha_j} + \frac{A_j V_j}{\alpha_j} \frac{\pi^2 \alpha_j (\alpha_j -1)}{6} T_j^{\alpha_j} \left(\frac{\mu_j}{T_j}\right)^{\alpha_j-2}\\ \nonumber
=\frac{A_j V_j}{\alpha_j} \left( x_j^{\alpha_j}  +  \frac{\pi^2 \alpha_j (\alpha_j -1)}{6} x_j^{\alpha_j-2}\right) T_j^{\alpha_j},
\end{gather}
\begin{gather}\label{EEE_SpL}
E_j=
\frac{A_j V_j}{\alpha_j+1} \left( x_j^{\alpha_j+1} + \frac{\pi^2 \alpha_j (\alpha_j + 1)}{6} x_j^{\alpha_j-1} \right) T_j^{\alpha_j+1}.
\end{gather}
Here $x_j=\mu_j/T_j$, $\alpha_j=D_j/d_j$.
Considering reservoirs' temperatures and chemical potentials as time functions $\mu_j=\mu_j(t),\ T_j=T_j(t),\ V_j=\mathrm{const}$ and using the procedure described in the main text (Eqs.~(4)-(5)), we get


\begin{gather}
\nonumber
\frac{dN_j/dt}{A_j V_j}=\left( x_j^{\alpha_j-1} + \frac{\pi^2}{6}(\alpha_j -1)(\alpha_j -2) x_j^{\alpha_j-3}\right) T_j^{\alpha_j} \dot x_j+ \\ 
+  \left( x_j^{\alpha_j} + \frac{\pi^2}{6}\alpha_j(\alpha_j -1) x_j^{\alpha_j-2}\right) T_j^{\alpha_j-1} \dot T_j,
\end{gather}

\begin{gather}
\frac{dE_j/dt}{A_j V_j}=\left( x_j^{\alpha_j} + \frac{\pi^2}{6}\alpha_j (\alpha_j -1) x_j^{\alpha_j-2}\right) T_j^{\alpha_j+1} \dot x_j + \\ \nonumber
+  \left( x_j^{\alpha_j+1} + \frac{\pi^2}{6}\alpha_j(\alpha_j +1) x_j^{\alpha_j-1}\right) T_j^{\alpha_j} \dot T_j.
\end{gather}
Thus, 
\begin{gather}\label{T_dyn_SpL}
\dot T_j = - \frac{-Q_{\alpha_j} T_j \overline{P}_j + Q_{\alpha_j-1}  \overline{J}_j }{ \tilde{A}_j V_j T_j^{\alpha_j} x_j^{2\alpha_j-4}\left(x_j^2-\cfrac{\pi^2}{6}\alpha_j(\alpha_j-1)\right)},
\end{gather}

\begin{gather}\label{x_dyn_SpL}
\dot x_j =  \frac{-Q_{\alpha_j+1} T_j \overline{P}_j +Q_{\alpha_j}  \overline{J}_j }{ \tilde{A}_j V_j T_j^{\alpha_j+1} x_j^{2\alpha_j-4}\left(x_j^2-\cfrac{\pi^2}{6}\alpha_j(\alpha_j-1)\right)}.
\end{gather}
Here $Q_{\alpha_j}=x_j^{\alpha_j} + \cfrac{\pi^2}{6}\alpha_j (\alpha_j -1) x_j^{\alpha_j-2}$, $\tilde{A}_j=A_j \pi^2 /3$.
These are Eqs.~(11) from the main text.

\begin{figure}
\begin{minipage}[h]{\linewidth}
\center{\includegraphics[width=\linewidth]{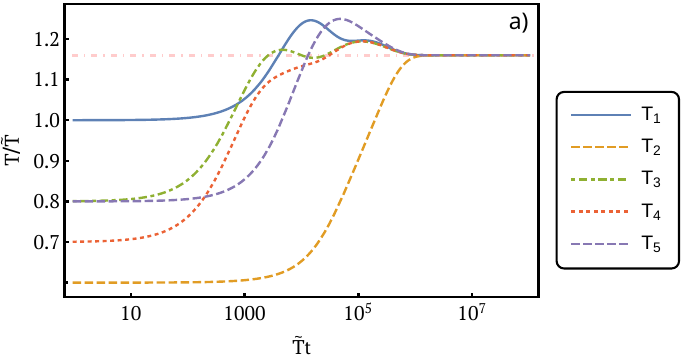}}
\end{minipage}
\begin{minipage}[h]{\linewidth}
\center{\includegraphics[width=\linewidth]{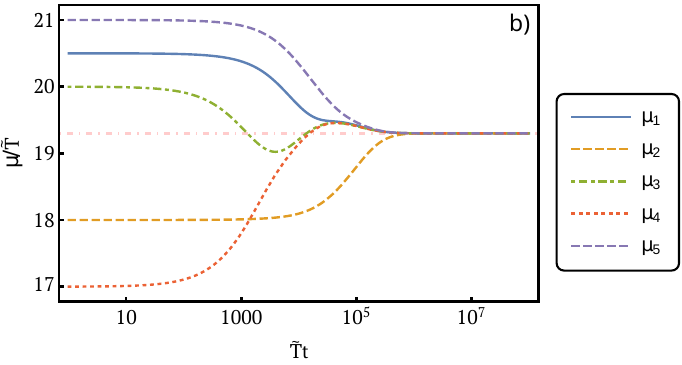}}
\end{minipage}
\begin{minipage}[h]{\linewidth}
\center{\includegraphics[width=\linewidth]{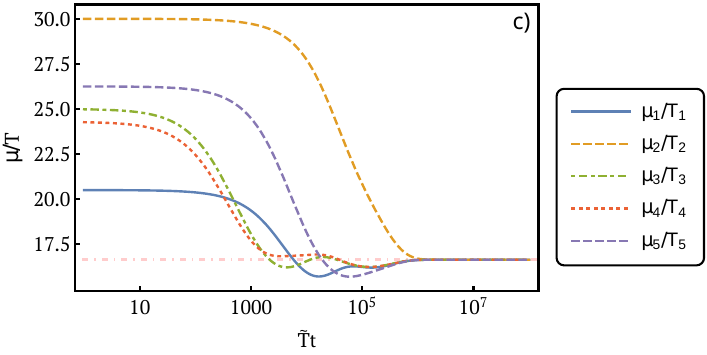}}
\end{minipage}
\caption{Dynamics of five reservoirs' temperatures and chemical potentials: $\omega_1/\mathrm{\tilde{T}}=18.5$, $\omega_2/\mathrm{\tilde{T}}=18.0$, $\omega_3/\mathrm{\tilde{T}}=20.5$, $\omega_4/\mathrm{\tilde{T}}=21.0$, $\omega_5/\mathrm{\tilde{T}}=17.0$, $\omega_6/\mathrm{\tilde{T}}=17.5$,
$T_1/\mathrm{\tilde{T}}=1.0$, $T_2/\mathrm{\tilde{T}}=0.6$, 
$T_3/\mathrm{\tilde{T}}=0.8$,
$T_4/\mathrm{\tilde{T}}=0.7$,
$T_5/\mathrm{\tilde{T}}=0.8$,
$\mu_1/\mathrm{\tilde{T}}=20.5$, $\mu_2/\mathrm{\tilde{T}}=18.0$, 
$\mu_3/\mathrm{\tilde{T}}=20.0$, 
$\mu_4/\mathrm{\tilde{T}}=17.0$, 
$\mu_5/\mathrm{\tilde{T}}=21.0$,
$\gamma_{1}(\omega)=10^{-4}(\omega/\mathrm{\tilde{T}})^{\alpha_{1}-1}$,
$\gamma_{2}(\omega)=10^{-5}(\omega/\mathrm{\tilde{T}})^{\alpha_{2}-1}$, 
$\gamma_{3}(\omega)=10^{-3}(\omega/\mathrm{\tilde{T}})^{\alpha_{3}-1}$, 
$\gamma_{4}(\omega)=0.5\cdot 10^{-3}(\omega/\mathrm{\tilde{T}})^{\alpha_{4}-1}$, 
$\gamma_{5}(\omega)=0.5\cdot 10^{-4}(\omega/\mathrm{\tilde{T}})^{\alpha_{5}-1}$, $\alpha_{j}=3/2$.}
\label{N5}
\end{figure}

Equilibrium temperature and chemical potential can be found via total energy and number of particles conservation 
\begin{gather}\label{Eq_T_mu_SpL}
    \sum_j N(T_{eq},\mu_{eq})=\sum_j N(T_j(0),\mu_j(0)), \\ \nonumber
    \sum_j E(T_{eq},\mu_{eq})=\sum_j E(T_j(0),\mu_j(0)).
\end{gather}

Using Eqs.~(\ref{NNN_SpL})-(\ref{EEE_SpL}), we get ($B_j=A_j V_j /\alpha_j$)
\begin{gather}
    \sum_{j=1}^n B_j \left( \mu_j^{\alpha_j}  +  \frac{\pi^2 \alpha_j (\alpha_j -1)}{6} \mu_j^{\alpha_j}\frac{T_j^2}{\mu_j^2}\right)=\\ \nonumber
    =\sum_{j=1}^n B_j \left( \mu_{eq}^{\alpha_j}  +  \frac{\pi^2 \alpha_j (\alpha_j -1)}{6} \mu_{eq}^{\alpha_j}\frac{T_{eq}^2}{\mu_{eq}^2}\right)
\end{gather}
\begin{gather}
    \sum_{j=1}^n B_j\frac{\alpha_j \mu_j}{\alpha_j+1} \left( \mu_j^{\alpha_j}  +  \frac{\pi^2 \alpha_j (\alpha_j +1)}{6} \mu_j^{\alpha_j}\frac{T_j^2}{\mu_j^2}\right)=\\ \nonumber
    =\sum_{j=1}^n B_j \frac{\alpha_j \mu_j}{\alpha_j+1} \left( \mu_{eq}^{\alpha_j}  +  \frac{\pi^2 \alpha_j (\alpha_j +1)}{6} \mu_{eq}^{\alpha_j}\frac{T_{eq}^2}{\mu_{eq}^2}\right)
\end{gather}

We can solve the obtained system of equations explicitly in the case of 2D reservoirs with quadratic dispersion ($\alpha_j=1$).
Let denote here $\langle O \rangle= \sum_j O_j B_j/\sum_j B_j$ (weighted value of $O_j$).
Then,
\begin{gather}
    \mu_{eq}= \langle \mu \rangle,\ \ \ \frac{\langle \mu^2 \rangle-\langle \mu \rangle^2}{\pi^2/3}+\langle T^2 \rangle=T_{eq}^2.\\ \nonumber
    \min_j\mu_j\le \mu_{eq}\le\max_j \mu_j,\\ \nonumber
    \min_j T_j^2+\frac{D_\mu}{\pi^2/3}\le T_{eq}^2\le \max_j T_j^2+\frac{D_\mu}{\pi^2/3}.
\end{gather}
Here $D_\mu= \langle \mu^2 \rangle - \langle \mu \rangle ^2\ge 0$ is analog to dispersion.
Hence, equilibrium chemical potential should always be between instant maximal and minimal chemical potentials.
On the other hand, equilibrium temperature has its minimal possible value, which is greater than the minimal initial temperature among reservoirs, and its maximal possible value, which is greater than the maximal initial temperature among reservoirs.

Despite the fact that quantitative estimations are done for the case of planar (2D) reservoirs, discussed behavior can be noticed in figures in the main text for 3D reservoirs.
Here we also present an example of temperatures' and chemical potentials' equalization in the case of five 3D reservoirs (see Fig.~\ref{N5}).
It is seen that the discussed properties of 2D reservoirs' equilibrium temperatures and chemical potentials also manifest here.

\subsection{Direct proof of SLoT fulfillment for considered reservoirs}\label{SpL_C}

In the quasi-stationary state of the OQS, $\dot{S}_S\approx 0$.
Thus, $\sum_{j=1}^n \overline{Y}_j/T_j \le 0$.
Strictly speaking, the fulfillment of the $\sum_{j=1}^n \overline{Y}_j/T_j \le 0$ condition follows from the Davies form of the used ME Eq.~(\ref{ME_SpL}) \cite{spohn1973irreversible,spohn1978entropy}.
However, we also provide independent proof to underline that the results achieved in the main text do not break SLoT.
In the main text we have used stationary flows in the following form:
\begin{gather}\label{J_stat_SpL}
    \overline{J}_j=\sum_\kappa \omega_\kappa \gamma_j (\omega_\kappa)(n_j(\omega_\kappa,\mu_j, T_j) - \tilde{n}(\omega_\kappa)),\\ \nonumber
    \overline{P}_j=\sum_\kappa \gamma_j (\omega_\kappa)(n_j(\omega_\kappa,\mu_j, T_j) - \tilde{n}(\omega_\kappa)), \\ \nonumber
    \overline{Y}_j=\overline{J}_j-\mu_j\overline{P}_j=\\ \nonumber
    =\sum_\kappa (\omega_\kappa - \mu_j)\gamma_j (\omega_\kappa)(n_j(\omega_\kappa,\mu_j, T_j) - \tilde{n}(\omega_\kappa)).
\end{gather}

We can rewrite these equations as follows:
\begin{gather}
    n_j(\omega_\kappa,\mu_j, T_j) - \tilde{n}(\omega_\kappa)=\\ \nonumber
    =n_j(\omega_\kappa,\mu_j, T_j) -\cfrac{\sum_{q=1}^n \gamma_q (\omega_\kappa) n_q(\omega_\kappa,\mu_q, T_q)}{\sum_{q=1}^n 
 \gamma_q (\omega_\kappa)}=\\ \nonumber
 =\frac{\sum_{q=1}^n 
 \gamma_q (\omega_\kappa)(n_j(\omega_\kappa,\mu_j, T_j) - n_q(\omega_\kappa,\mu_q, T_q))}{\sum_{q=1}^n 
 \gamma_q (\omega_\kappa)}=\\ \nonumber
 =-2\sum_{q=1}^n \cfrac{\gamma_q (\omega_\kappa)}{\Sigma_\kappa}
 \frac{\mathrm{sinh}\left(\mathrm{v}_{\kappa j}-\mathrm{v}_{\kappa q}\right)}{\mathrm{cosh}\left(\mathrm{v}_{\kappa j}\right) \mathrm{cosh}\left(\mathrm{v}_{\kappa q}\right)}.
\end{gather}
Here $\Sigma_\kappa=\sum_{q=1}^n \gamma_q(\omega_\kappa)$, $\mathrm{v}_{\kappa j}=\cfrac{\omega_\kappa - \mu_j}{T_j}$.
Then
\begin{gather}
    \overline{J}_j=-2\sum_\kappa \sum_{q=1}^n \omega_\kappa \Gamma_{jq\kappa}\frac{\mathrm{sinh}\left(\mathrm{v}_{\kappa j}-\mathrm{v}_{\kappa q}\right)}{\mathrm{cosh}\left(\mathrm{v}_{\kappa j}\right) \mathrm{cosh}\left(\mathrm{v}_{\kappa q}\right)}, \\ \nonumber
    \overline{P}_j=-2\sum_\kappa \sum_{q=1}^n \Gamma_{jq\kappa}\frac{\mathrm{sinh}\left(\mathrm{v}_{\kappa j}-\mathrm{v}_{\kappa q}\right)}{\mathrm{cosh}\left(\mathrm{v}_{\kappa j}\right) \mathrm{cosh}\left(\mathrm{v}_{\kappa q}\right)}, \\ \nonumber
    \overline{Y}_j=-2\sum_\kappa \sum_{q=1}^n T_j \mathrm{v}_{\kappa j} \Gamma_{jq\kappa}\frac{\mathrm{sinh}\left(\mathrm{v}_{\kappa j}-\mathrm{v}_{\kappa q}\right)}{\mathrm{cosh}\left(\mathrm{v}_{\kappa j}\right) \mathrm{cosh}\left(\mathrm{v}_{\kappa q}\right)}.
\end{gather}
Here $\Gamma_{jq\kappa}=\gamma_j(\omega_\kappa)\gamma_q (\omega_\kappa)/\Sigma_\kappa$.
It is seen that $\Gamma_{jq\kappa}=\Gamma_{qj\kappa}\ge 0$.
Using the achieved form of heat flows, we get
\begin{gather}
    \sum_{j=1}^n \frac{\overline{Y}_j}{T_j}
    =-2\sum_\kappa \sum_{q=1}^n \sum_{j=1}^n \mathrm{v}_{\kappa j} \Gamma_{jq\kappa}\frac{\mathrm{sinh}\left(\mathrm{v}_{\kappa j}-\mathrm{v}_{\kappa q}\right)}{\mathrm{cosh}\left(\mathrm{v}_{\kappa j}\right) \mathrm{cosh}\left(\mathrm{v}_{\kappa q}\right)}.
\end{gather}

Summing this equation with a similar one that has inverted indices $q$ and $j$, we obtain
\begin{gather}
\nonumber
    \sum_{j=1}^n \frac{\overline{Y}_j}{T_j}
    =-\sum_\kappa \sum_{q=1}^n \sum_{j=1}^n \mathrm{v}_{\kappa j} \Gamma_{jq\kappa}\frac{\mathrm{sinh}\left(\mathrm{v}_{\kappa j}-\mathrm{v}_{\kappa q}\right)}{\mathrm{cosh}\left(\mathrm{v}_{\kappa j}\right) \mathrm{cosh}\left(\mathrm{v}_{\kappa q}\right)}\\ \label{Uss_Eqq_SpL}
    -\sum_\kappa \sum_{q=1}^n \sum_{j=1}^n \mathrm{v}_{\kappa q} \Gamma_{qj\kappa}\frac{\mathrm{sinh}\left(\mathrm{v}_{\kappa q}-\mathrm{v}_{\kappa j}\right)}{\mathrm{cosh}\left(\mathrm{v}_{\kappa q}\right) \mathrm{cosh}\left(\mathrm{v}_{\kappa j}\right)}\\ \nonumber
    =-\sum_\kappa \sum_{q=1}^n \sum_{j=1}^n (\mathrm{v}_{\kappa j}-\mathrm{v}_{\kappa q}) \Gamma_{jq\kappa}\frac{\mathrm{sinh}\left(\mathrm{v}_{\kappa j}-\mathrm{v}_{\kappa q}\right)}{\mathrm{cosh}\left(\mathrm{v}_{\kappa j}\right) \mathrm{cosh}\left(\mathrm{v}_{\kappa q}\right)}.
\end{gather}

In Eq.~(\ref{Uss_Eqq_SpL}) each term under the sum is equal to or greater than zero.
Thus, $\sum_{j=1}^n \overline{Y}_j/T_j \le 0$ and SLoT is fulfilled.

\subsection{The equilibrium temperature and chemical potential}\label{SpL_Equlib}

In equilibrium state of Eq.~(\ref{T_dyn_SpL})-(\ref{x_dyn_SpL}) $\overline{P}_j=0$, $\overline{J}_j=0$ as for the considered case $\mu_j/T_j\gg 1$ matrix of this system can be inverted.
Thus, in equilibrium state $\overline{Y}_j=0$.

The sum in Eq.~(\ref{Uss_Eqq_SpL}) contains terms of the same sign.
Hence, if the total sum equals zero, each term in the sum equals zero.
Hence, for any $\kappa,j,q$ we have $(\omega_\kappa-\mu_j)/T_j=(\omega_\kappa-\mu_q)/T_q$.

This is satisfied independently on $\omega_\kappa$, if all reservoirs' temperatures and chemical potentials are equal.
However, it is needed to be shown that this is the only one possibility to get all $\overline{Y}_j=0$.
We show this by contradiction.

Let suppose that reservoirs $j$ and $q$ have different temperatures in equilibrium state.
Then for any $ \kappa$ holds $ \omega_\kappa=-(\mu_j T_q-\mu_q T_j)/(T_j-T_q)$.
This is impossible, if there are at least two different $\omega_\kappa$ in the spectrum of the OQS.

On the other hand, if equilibrium temperatures of $j$-th and $q$-th reservoirs are the same, equilibrium chemical potentials are also equal $\mu_j=\mu_q$.
Thus, if there are at least two different $\omega_\kappa$ in the OQS, from $\overline{P}_j=0$, $\overline{J}_j=0$ it follows that all reservoirs' temperatures are equal $T_1=\cdots =T_n$, and all reservoirs' chemical potentials are equal too $\mu_1=\cdots=\mu_n$.


\end{document}